\documentclass[aps,pra,superscriptaddress,twocolumn,floatfix]{revtex4-1}
\usepackage{graphicx}
\usepackage{amssymb}
\usepackage{amsfonts}
\usepackage{amsmath}
\usepackage{epstopdf}

\begin{document}

\title{Superconducting and ferromagnetic phase diagram of UCoGe probed by thermal expansion}

\author{A. M. Nikitin} \email{a.nikitin@uva.nl}\affiliation{Van der Waals - Zeeman Institute, University of Amsterdam, Science Park 904, 1098 XH Amsterdam, The Netherlands}
\author{J. Geldhof} \affiliation{Van der Waals - Zeeman Institute, University of Amsterdam, Science Park 904, 1098 XH Amsterdam, The Netherlands}
\author{Y. K. Huang} \affiliation{Van der Waals - Zeeman Institute, University of Amsterdam, Science Park 904, 1098 XH Amsterdam, The Netherlands}
\author{D. Aoki} \affiliation{IMR, Tohoku University, Oarai, Ibaraki 311-1313, Japan} \affiliation{Universit\'{e} Grenoble Alpes, CEA, INAC-PHELIQS, F-38000 Grenoble, France}
\author{A. de Visser} \email{a.devisser@uva.nl} \affiliation{Van der Waals - Zeeman Institute, University of Amsterdam, Science Park 904, 1098 XH Amsterdam, The Netherlands}

\date{\today}

\begin{abstract}
We report thermal expansion measurements on a single crystal of the superconducting ferromagnet UCoGe for magnetic fields applied along the main orthorhombic axes. The thermal expansion cell was mounted on a piezo-electric rotator in order to fine-tune the magnetic field angle. The superconducting and magnetic phase diagram has been determined. With our bulk technique we confirm the $S$-shape of the upper-critical field, $B_{c2}$, for $B \parallel b$ and reinforcement of superconductivity above 6~T. At the same time the Curie point shifts towards lower temperatures on increasing the field along the \emph{b}-axis. This lends further support to theoretical proposals of spin-fluctuation mediated reinforcement of superconductivity for $B \parallel b$.
\end{abstract}

\date{\today}


\maketitle

\section{INTRODUCTION}

The quest for superconducting ferromagnets (SCFMs) goes back to the pioneering work of Ginzburg~\cite{Ginzburg1957}. Later, in the 1970s, it was predicted, on theoretical grounds, that weak itinerant ferromagnets could exhibit $p$-wave equal-spin-pairing superconductivity~\cite{FayAppel}. Here, the superconducting state is mediated by the exchange of longitudinal spin fluctuations, rather than by phonons as in the standard Bardeen, Cooper, Schrieffer (BCS) scenario. The experimental discovery of superconductivity in the itinerant ferromagnets UGe$_2$ (under pressure)~\cite{Saxena2000}, URhGe~\cite{Aoki2001} and UCoGe~\cite{Huy2007}, opened up the opportunity to investigate the coexistence of superconductivity and ferromagnetism and their interplay meticulously. In SCFMs the superconducting transition temperature $T_{sc}$ is smaller than the Curie temperature $ T_C$. The uranium \textit{5f}-electrons are responsible for both ferromagnetic order and superconductivity. In the past decade ample evidence has been provided that spin-fluctuations in the proximity to a magnetic quantum critical point provide the attractive pairing interaction for odd-parity superconductivity (for recent reviews see Refs.~\cite{AokiJPSJ2014,Huxley2015}). Although these SCFMs share common features, they are also quite distinct, notably as regards the phase diagrams in the magnetic-field $-$ temperature and the pressure $-$ temperature plane.

UCoGe crystallizes in the orthorhombic TiNiSi structure with space group $P_{nma}$~\cite{Canepa}. Superconductivity at $T_{sc} = 0.6$~K and itinerant ferromagnetism at $T_C = 3.0$~K was first observed for polycrystalline samples~\cite{Huy2007}. The coexistence of ferromagnetism and superconductivity on the microscopic scale was demonstrated by muon spin rotation and relaxation ($\mu$SR)~\cite{VissermRS} and $^{59}$Co nuclear quadrupole resonance (NQR)~\cite{Ohta2010} experiments. Magnetization measurements on single-crystals showed that UCoGe is an uniaxial ferromagnet with a spontaneous magnetic moment, $m_0$ = 0.07 $\mu_B$ per U-atom, that points along the $c$-axis~\cite{Huy2008}. Superconductivity shows a strongly anisotropic response to a magnetic field. For $B$ parallel to the easy-magnetization axis ($c$-axis) superconductivity is suppressed in 0.5~T ($T \rightarrow 0$). On the other hand, for $B$ perpendicular to the easy-magnetization axis, the upper critical field, $B_{c2} (0)$, attains extremely large values and largely exceeds the Pauli limit for spin-singlet superconductivity~\cite{Huy2008,AokiJPSJ2009}. Moreover, $B_{c2}(T)$ measured for $B \parallel b$-axis~\cite{AokiJPSJ2009} shows a striking $S$-shape which signals reinforced superconductivity in fields exceeding 6~T. The pronounced anisotropy of $B_{c2}$ is arguably coupled to spin-fluctuation mediated pairing: for $B \parallel c$ the ferromagnetic fluctuations are suppressed and superconductivity vanishes, while for $B \perp c$ spin fluctuations are robust and superconductivity persists up to high fields. Solid experimental evidence on the microscopic level for this scenario has been provided by $^{59}$Co nuclear magnetic resonance (NMR)~\cite{Ihara2010,Hattori2012} and inelastic neutron scattering~\cite{Stock2011} experiments. The superconducting phase diagram was qualitatively explained by a microscopic theory employing the coupling between the electrons by means of magnetization fluctuations in ferromagnetic metals~\cite{Mineev2014}. A related approach was based on the  Eliashberg theory by taking into account the Dzyaloshinskii-Moriya interaction arising from the zigzag chain crystal structure~\cite{Tada2016}. Here the $S$-shaped $B_{c2}$-curve is qualitatively explained as a result of the enhancement of the spin fluctuations due to the decrease of the Curie temperature when the field $B \parallel b$ is increased.

In this paper we report the ferromagnetic and superconducting phase diagram in the $B-T$ plane of UCoGe obtained by thermal expansion measurements in fixed magnetic fields applied along the $a$-, $b$- and $c$-axis. Hitherto, the phase diagram was mainly studied using transport experiments~\cite{AokiJPSJ2009}. Its determination by thermal expansion has the advantage that it involves a thermodynamic bulk probe. Moreover, since the phase diagram depends sensitively on the alignment of the field with respect to the $a$- or $b$-axis~\cite{AokiJPSJ2009}, we mounted our thermal expansion cell on a piezo-electric rotator to enable tuning of the field-angle. We observe that the Curie point for $B \parallel b$ shifts gradually to lower temperatures and we present bulk-sensitive evidence for the $S$-shaped $B_{c2}$-curve for the same field orientation. Our results lend further support to the theoretical proposal of spin-fluctuation mediated enforcement of superconductivity in a magnetic field ($B \parallel b$).

\section{EXPERIMENTAL}

Single crystals of UCoGe were prepared in a tri-arc furnace (crystal~1) and in a tetra-arc furnace (crystal~2) using the Czochralski technique. Crystal~1, with residual resistance ratio RRR = 30,  was cut from the grown crystal boule by means of spark erosion into a bar shape with dimensions $a \times b \times c = 1.0 \times 5.0 \times 1.1$~mm$^{3}$. Sample~2 has RRR = 40 and was cut into a cube-like shape with dimensions $a \times b \times c = 1.4 \times 1.1 \times 1.0$~mm$^{3}$. Here RRR is defined as $R$(300K)/$R$(0K), where $R$(0K) is obtained by extrapolating the normal state resistance $R_n(T)$ to 0~K. The uncertainty in the alignment of the main crystallographic axes with the cut planes is typically 2$^{\circ}$. Additional information about the crystal synthesis, annealing procedure and characterization can be found in Refs.~\onlinecite{Huy2009JMMM,AokiJPSJ2012}.

The coefficient of linear thermal expansion, $\alpha = L^{-1}(dL/dT)$, with $L$ the sample length, was measured using a three-terminal parallel-plate capacitance method. The home-built sensitive dilatometer~\cite{Nikitinthesis} was based on the design reported in Ref.~\onlinecite{Schmiedeshoff2006}. The sensitivity of the thermal expansion cell amounts to 0.03 \AA. The dilatometer was used in two configurations: longitudinal, \textit{i.e.} with the field along the dilatation direction $B \parallel \Delta L$, and transversal with $B \perp \Delta L$. In order to tune the magnetic field-angle with respect to the crystal axes in the transverse configuration, we have implemented an \textit{in situ} rotation mechanism~\cite{Nikitinthesis}. The thermal expansion cell was mounted on a piezo-electric rotator (Attocube ANRv220/RES) with help of a horizontal-to-vertical motion gear set with a gear ratio of 1 to 3, which allowed us to reach an angular resolution of 0.05$^{\circ}$. In this configuration the magnetic field is always perpendicular to the fixed dilatation direction when the cell is rotated. The standard field-angle reader of the rotator, the positioner, was calibrated at low temperatures using a miniature Hall-probe (Arepoc company) mounted on the main body of the cell. We remark that since we can rotate over one axis only, a possible remaining misorientation of $\sim 2^{\circ}$ due to orienting and cutting the crystal cannot be avoided. The dilatometer and rotator were attached to the cold plate of a Helium-3 refrigerator for temperatures in the range $T=0.24 - 10$~K (Heliox, Oxford Instruments) and magnetic fields up to 14~T, and to the cold finger of a dilution refrigerator (Kelvinox, Oxford Instruments) for $T = 0.03 - 1$~K and $B$ up to 17~T. An additional heater and thermometer were mounted, thermally anchored to the thermal expansion cell, for the step-wise heating method to measure $\alpha (T)$.

\section{RESULTS}

\subsection{Thermal expansion in zero field}

\begin{figure}
\includegraphics[width=8cm]{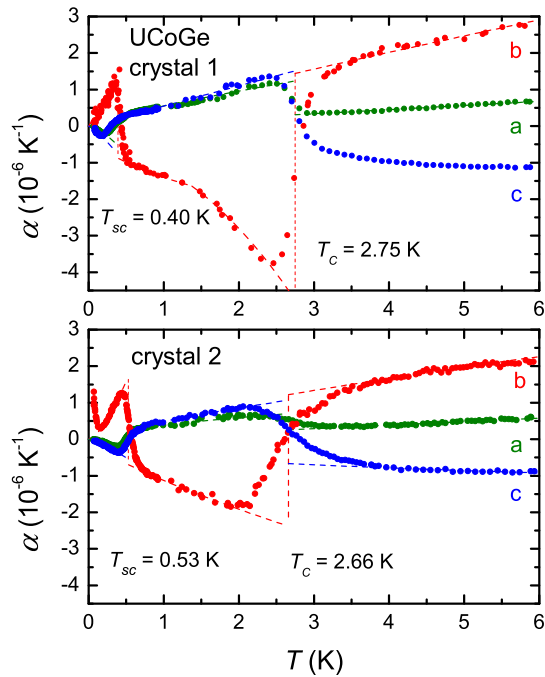}
\caption{Temperature variation of the coefficient of linear thermal expansion of UCoGe along the orthorhombic \emph{a}-, \emph{b}- and \emph{c}-axis, as indicated, for crystal~1 (top panel) and crystal~2 (bottom panel). The data for $\alpha_b$ in the top panel are taken from Ref~\onlinecite{Gasparini2010}. The dashed lines indicate idealized stepped phase transitions based on an equal-length construction (see text).}
\label{figure:zfte}
\end{figure}

The thermal expansion coefficient, $\alpha_{i}(T)$, where $i$ refers to the crystal axis $a$, $b$ or $c$, of UCoGe crystal~1 and crystal~2, measured in the temperature range 0.05-6.0~K, is shown in Fig.~\ref{figure:zfte}. The data for $\alpha_b$ in the upper panel are taken from Ref.~\onlinecite{Gasparini2010}, since the sample length of  crystal~1 along the $b$-axis (5~mm) is too large to fit in the dilatometer. The $\alpha_b$ data were measured on a different crystal with RRR = 30 prepared from the same batch. Overall $\alpha_{i}(T)$ of both crystals is very similar and in good agreement with the data reported in Ref.~\onlinecite{Gasparini2010}. The thermal expansion coefficient is strongly anisotropic in the paramagnetic state. Below $T_C$, $\alpha_a(T)$ and $\alpha_c(T)$ behave similarly, while the most pronounced and opposite variation is found for $\alpha_b(T)$. The steps in $\alpha_{i}(T)$ at $T_C$ and $T_{sc}$ are relatively broad. For the ferromagnetic transition they have a positive sign (when cooling) along the $a$- and $c$-axis, and a negative sign along the $b$-axis. At the superconducting phase transition the signs are reversed. The dashed lines in Fig.~\ref{figure:zfte} represent idealized transitions, based on an equal-length construction, with $T_{sc}$ = 0.40~K and $T_C$ = 2.75~K for crystal~1 and $T_{sc}$ = 0.53~K and $T_C$ = 2.66~K for crystal~2. The construction implies an overall equal-length change is imposed for the broadened and the idealized contributions when integrating $\alpha_{i}(T)$ with respect to the background.

\begin{figure}
\includegraphics[width=8cm]{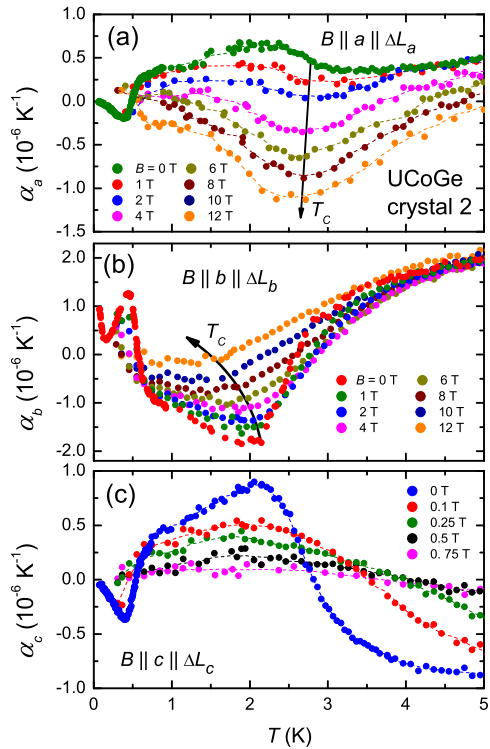}
\caption{Coefficient of linear thermal expansion $\alpha _i$, where $i$ is $a$, $b$ or $c$, of UCoGe crystal~2 in longitudinal magnetic fields $B \parallel a,b,c$ as indicated. Arrows indicate the field variation of $T_{C}$ for $B \parallel a$ and $B \parallel b$.}
\label{figure:lfte}
\end{figure}

The different values of both $T_{sc}$ and $T_{C}$ for the two crystals demonstrate the intimate interplay between ferromagnetism and superconductivity in UCoGe. For the crystal with lower $T_{C}$, the superconducting transition temperature is slightly higher, which is in-line with the enhancement of superconductivity when the magnetic quantum critical point is approached~\cite{Slooten2009}. We remark that the data for crystal~2 show a pronounced upturn in $\alpha_b$ at very low temperatures ($T < 0.15$~K). This anomalous behavior is only observed for $\alpha_b$, and not for $\alpha_a$ and $\alpha_c$. Its origin is not understood as will be discussed in section IV-C. In the following sections we present the thermal expansion in magnetic field in the longitudinal and transverse geometry for crystal~2.

\subsection{Longitudinal thermal expansion in magnetic field}

\begin{figure}
\includegraphics[width=8cm]{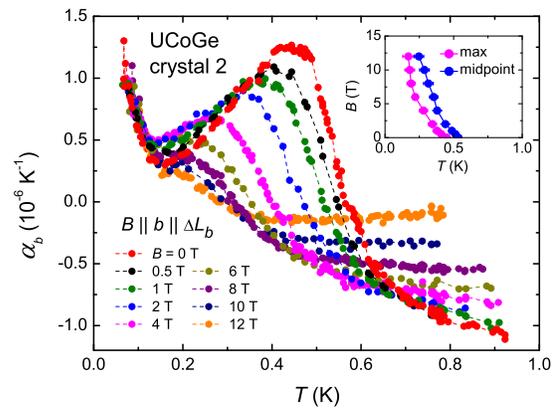}
\caption{Coefficient of linear thermal expansion $\alpha _b$ of UCoGe crystal~2 in magnetic fields $B \parallel \Delta L_b$ up to 12~T. Inset: \emph{B-T} phase diagram. $T_{sc}$ obtained by tracking the maximum in $\alpha _b (T)$ (magenta points) and by the midpoints of the transition (blue points).}
\label{figure:lfte2}
\end{figure}

In Fig.~\ref{figure:lfte} we present the longitudinal thermal expansion of crystal~2 around the Curie point in fields up to 12~T ($B \parallel \Delta L_i$). In this configuration the thermal expansion cell is directly attached to the cold plate of the Heliox or cold finger of the Kelvinox, \textit{i.e.} without rotator, thus tuning of the field-angle is not possible. The response to the magnetic field is different for each direction. For $B \parallel \Delta L_c$ the ferromagnetic phase transition smears out rapidly. In 0.75~T $\alpha_c (T)$ is quasi temperature independent and close to zero. This is expected since the field is parallel to the ordered moment $m_0$ and the phase transition becomes a cross-over phenomenon. For the other two directions $B \perp m_0$ and the Curie point remains clearly visible in the data. In the case of $B \parallel \Delta L_a$ the magnetic contribution to $\alpha_a$ changes from positive to negative between 2~T and 4~T. In higher fields the magnetic component grows further, while the transition broadens. The Curie temperature, which we identify by the minimum in $\alpha_a (T)$ at higher fields, is only weakly field dependent. For $B \parallel b \parallel \Delta L$ the magnetic contribution becomes weaker as the field grows, and $T_{C}$ shifts towards lower temperatures. Comparing the data in field for crystal~2 with those reported in Ref.~\onlinecite{Gasparini2010} we find a good agreement for $B \parallel \Delta L_c$. For the other two directions the literature data show a more pronounced magnetic contribution in the field. This we attribute to the more developed ferromagnetic phase in the crystal measured in Ref.~\onlinecite{Gasparini2010} (just like for crystal~1 in zero-field, as can be seen in Fig.~1). The longitudinal thermal expansion around the superconducting transition was measured in field only for $\alpha_b$ and is reported in Fig.~3. The superconducting transition observed at $T_{sc} = 0.53$~K in zero field shifts to lower temperatures with increasing field and remains clearly visible up to 12~T (see Fig.~\ref{figure:lfte2}). The data in Fig.~\ref{figure:lfte2} also show the upturn in $\alpha_b (T)$ below 0.15~K persists in the magnetic field.

\subsection{Transverse thermal expansion in magnetic field}

\begin{figure}
\includegraphics[width=8cm]{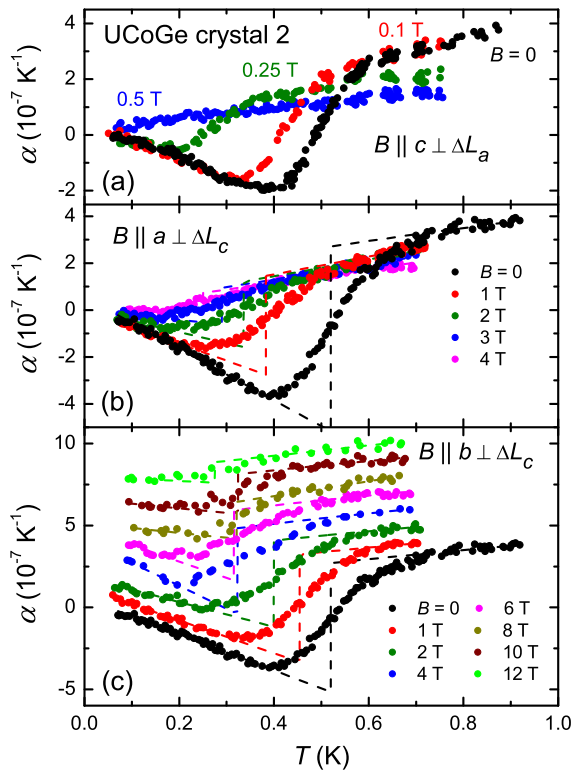}
\caption{Coefficient of linear thermal expansion of UCoGe crystal~2 in transverse magnetic fields. Panel (a): For  $\Delta L _a$ and $B \parallel c$ (up to 0.5~T as indicated). Panel (b): For $\Delta L _c$ and $B \parallel a$ (up to 4~T). Panel (c): For $\Delta L _c$ and $B \parallel b$ (up to 12~T). In panel (c) the curves in field are shifted upwards for clarity. The dashed lines indicate idealized step-like transitions.}
\label{figure:tfte1}
\end{figure}

The upper-critical field, $B_{c2}$, of UCoGe is strongly anisotropic~\cite{Huy2008}. For $B \parallel c$ $B_{c2}(0)= 0.5$~T ($T \rightarrow 0$), but when the field is precisely aligned along the $a$- or $b$-axis $B_{c2}(0)$ attains extremely large values with 16~T for $B \parallel b$ and close to 25~T for $B \parallel a$~\cite{AokiJPSJ2009}. A field-tilt of a few degrees away from the $a$- or $b$-axis results in a dramatic reduction of $B_{c2}(0)$. In order to fine-tune the field angle, the dilatometer was mounted on the rotator and the transverse thermal expansion was measured. The suppression of the superconducting state in the case $B \parallel c$ was measured for $B \parallel c \perp \Delta L_a$. The data are shown in Fig.~\ref{figure:tfte1}~(a). Superconductivity is gradually depressed towards lower temperatures with increasing field and is no longer observed at $B = 0.5$~T. The suppression of the superconducting state for $B \parallel a (b)$ was measured for $B \parallel a (b) \perp \Delta L_c$. To achieve an optimal alignment of the field along the $a$- and $b$-axis we have used the following strategy. $\alpha_c (T)$ around the superconducting transition was measured in a field of 1~T, then the dilatometer was rotated over typically 0.5$^{\circ}$ and $\alpha_c (T)$ was measured again. After obtaining several data sets in this way we selected the optimal orientation $B \parallel a (b)$ as the one in which $\alpha_c (T)$ shows the highest $T_{sc}$. The resulting curves are presented in Fig.~\ref{figure:tfte1}~(b) and (c). For $B \parallel a$ the superconducting transition broadens rapidly and we can follow it only to 4~T. For $B \parallel b$ we find a very different behavior. Superconductivity is first depressed but then stabilizes and is reinforced. In fact the superconducting phase transition is detectable up to 12~T. For larger fields the noise level becomes higher than the idealized step in the linear thermal expansion coefficient at $T_{sc}$.

\begin{figure}
\includegraphics[width=8cm]{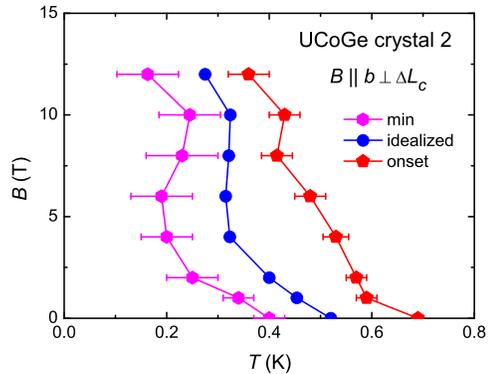}
\caption{\emph{B-T} phase diagram. $T_{sc}$ obtained from transverse configuration $B \parallel b \perp \Delta L_c$ by tracking the onset in $\alpha (T)$ data (red points), idealized transition (blue points) and minimum of the curve (magenta points).}
\label{figure:tfte2}
\end{figure}

In Fig.~\ref{figure:tfte2} we show the resulting $B-T$ phase diagram obtained by tracking $T_{sc}$ via three methods: the onset of the superconducting transition, the step in the idealized transition and the temperature of the local minimum in $\alpha_c (T)$. It is clear from Fig.~\ref{figure:tfte2} that the upper-critical field for $B \parallel b$ displays an  \emph{S}-shape curve with reinforcement of $B_{c2}$ for fields above 6~T.

\section{ANALYSIS AND DISCUSSION}

\subsection{Uniaxial and hydrostatic pressure dependence of critical temperatures}

\begin{table*}[t]
\caption{Idealized steps in $\alpha_i$ at the ferromagnetic (FM) and superconducting (SC) phase transitions for $i$ = $a$, $b$ and $c$ and the volume effect for UCoGe crystal~2 in zero field. The uniaxial and hydrostatic pressure dependencies of $T_C$ and $T_{sc}$ are calculated using the Ehrenfest relation (see text).} 
\centering 
\begin{tabular}{c c c c c}
\hline\hline
 & \emph{a} & \emph{b} & \emph{c} & \emph{volume} \\ [1.0ex] 
\hline
$\Delta \alpha_{FM}$ ($10^{-7}$/K) & 3.3 $\pm$ 0.5 & -33.2 $\mp$ 2.3 & 16.7 $\pm$ 1.1 & -13.2 $\mp$ 2.0 \\ [1.0ex]
$\partial T_C / \partial p_i$ (K/kbar) & 0.06 $\pm$ 0.01 & -0.63 $\mp$ 0.04 & 0.32 $\pm$ 0.02 & -0.25 $\mp$ 0.04 \\ [1.0ex]
$\Delta \alpha_{SC}$ ($10^{-7}$/K) & -4.9 $\mp$ 0.3 & 21.1 $\pm$ 2.3 & -7.5 $\mp$ 0.5 & 8.7 $\pm$ 0.9 \\ [1.0ex]
$\partial T_{sc} / \partial p_i$ (K/kbar) & -0.03 $\mp$ 0.001 & 0.15 $\pm$ 0.01 & -0.05 $\mp$ 0.003 & 0.07 $\pm$ 0.01 \\ [1.0ex] 
\hline\hline
\end{tabular}
\label{table:uniaxial} 
\end{table*}

The uniaxial and hydrostatic pressure dependence of $T_C$ and $T_{sc}$ of crystal 2 have been determined with help of the Ehrenfest relation. For a second-order phase transition $\frac{\partial T_{C,sc}}{\partial p_i} = \frac{V_m \Delta \alpha_i}{\Delta (c/T)}$, where the subscript $i$ refers to the orthorhombic axis, $V_m = 3.14 \times 10^{-5}$ m$^3$/mol is the molar volume and $\Delta (c/T)$ is the step in the specific heat divided by temperature at the transition. The specific heat data we use for the Ehrenfest analysis are reported in Fig.~\ref{figure:Gruneisen} and we obtain the idealized steps $\Delta (c/T)_{FM}$ = 16.6 mJ/molK$^2$ and $\Delta (c/T)_{sc}$ = 43.6 mJ/molK$^2$. We remark that these data were obtained on a different UCoGe crystal~\cite{AokiJPSJ2012} cut from the same batch as crystal~2. The step sizes in $\alpha_i$ and the resulting uniaxial and hydrostatic pressure dependencies are given in Table~\ref{table:uniaxial}. The largest uniaxial pressure effect is along the $b$-axis ($p_b$) for both the ferromagnetic and superconducting phase transitions. For $p_a$ and $p_c$ the effect is smaller and the sign is reversed compared to $p_b$. The calculated hydrostatic pressure variations amount to $\partial T_{C}/\partial p = -0.25$~K/kbar and $\partial T_{sc}/\partial p$ = 0.07 K/kbar. These values should be compared to $\partial T_{C}/\partial p = -0.79$~K/kbar and $\partial T_{sc}/\partial p$ = 0.10 K/kbar calculated with the Ehrenfest relation on a crystal comparable to crystal~1 as reported in Ref.~\onlinecite{Gasparini2010}. Our calculated values for the data shown here are close to the $\partial T_{C}/\partial p = -0.21$~K/kbar and $\partial T_{SC}/\partial p$ = 0.03~K/kbar values extracted from pressure dependent experiments~\cite{Hassinger2008,Slooten2009,BastienPRB16}.

\subsection{Gr{\"u}neisen analysis}

\begin{figure}
\includegraphics[width=8cm]{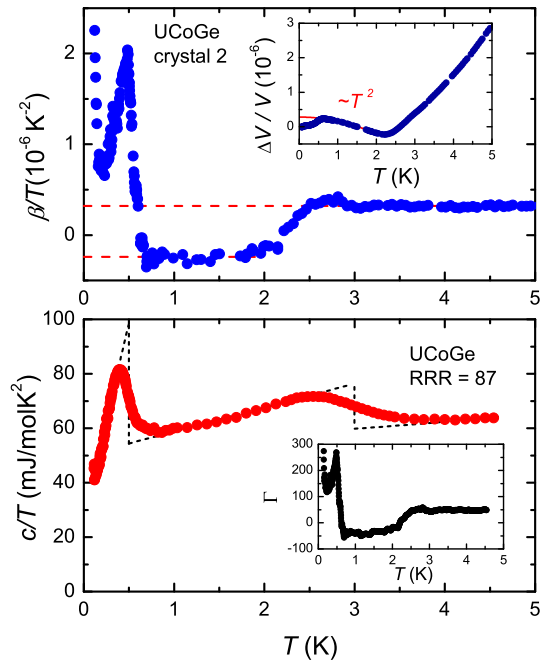}
\caption{Top panel: Coefficient of the volumetric thermal expansion divided by temperature, $\beta /T$, as a function of temperature of UCoGe (crystal~2). Inset: Temperature variation of the relative volume change $\Delta V/V$ of the same sample. Bottom panel: Specific heat divided by temperature, $c/T$, as a function of temperature of UCoGe (data taken from Ref.~\cite{AokiJPSJ2012}). Inset: Temperature variation of the Gr{\"u}neisen parameter, $\Gamma$.}
\label{figure:Gruneisen}
\end{figure}

In the top panel of Fig.~\ref{figure:Gruneisen} we present the volumetric thermal expansion coefficient divided by temperature, $\beta/T$, of UCoGe crystal~2, whereby $\beta = \alpha_a + \alpha_b + \alpha_c$. The ferromagnetic ordering results in a broad drop of $\beta/T$ below $T_C$, while the superconducting transition results in a sharp peak. In the paramagnetic phase, $\beta$ follows a linear temperature variation $\beta_{P} = a_{P} T $ ($T \leq 5$~K) with $a_{P} = 3.2 \times 10^{-7}$~K$^{-2}$. Also in the ferromagnetic phase, $\beta/T$ attains a constant value with $a_{FM} = -2.4 \times 10^{-7}$ K$^{-2}$. The relative volume change due to the ferromagnetic ordering and superconductivity is obtained by integrating $\beta (T)$ versus $T$ and is shown in the inset to the top panel of Fig.~\ref{figure:Gruneisen}. $\Delta V/V$ grows quadratically below the Curie point and decreases below 0.6~K due to superconductivity with $\Delta V/V = - 2.8 \times 10^{-7}$ for $T \rightarrow 0$. The latter value gives the spontaneous magnetostriction of the superconducting state and agrees well with the value previously obtained on UCoGe~\cite{Gasparini2010} and other heavy-fermions superconductors such as URu$_2$Si$_2$~\cite{vanDijk1995} and UPt$_3$~\cite{vanDijk1993}.

The effective Gr{\"u}neisen parameter $\Gamma$ is determined as $\Gamma (T) = \frac{V_m}{\kappa_T} \frac{\beta (T)}{c(T)}$, where $\kappa_T$ is the isothermal compressibility and $\beta (T)$ the volume expansion. For UCoGe $\kappa_T$ = 0.324~Mbar$^{-1}$, which was determined from the sum of the measured linear compressibilities along the \emph{a}-, \emph{b}- and \emph{c}-axis~\cite{Adamska2010,Maskova2012}. The resulting temperature dependence of the Gr{\"u}neisen parameter is shown in the inset to the bottom panel of Fig. \ref{figure:Gruneisen}. Above the Curie point, $\Gamma$ = 50. Upon cooling $\Gamma$ drops to a value of about $-50$ due to the ferromagnetic ordering and then dramatically increases to a value of about 300 due to the SC. In general, it is observed that Gr{\"u}neisen parameter increases rapidly with decreasing temperature as the heavy-fermion state stabilizes~\cite{Visser1990,vanDijk1995}. It should be noted that the negative Gr{\"u}neisen parameter for $T_{sc} < T < T_{C}$ implies the positive pressure derivative of the entropy ($dS/dp > 0$). This is in agreement with the collapse of $T_{C}$, applying the pressure. The large Gr{\"u}neisen parameter indicates that UCoGe is close to the quantum critical region. Indeed, non-Fermi liquid behaviour is observed in a wide pressure range near $p_c$~\cite{BastienPRB16}. The large Gr{\"u}neisen parameter is in contrast with that reported for another ferromagnetic superconductor, URhGe~\cite{Aoki2011573}.

\subsection{Phase diagram}

\begin{figure}
\includegraphics[width=8cm]{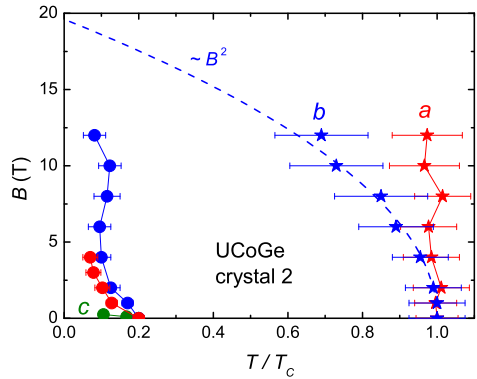}
\caption{Superconducting (solid circles) and ferromagnetic (stars) $B-T$ phase diagram of UCoGe (crystal 2) plotted as a function of the reduced temperature $T/T_C$, where $T_C$ is the Curie temperature. $T_{sc}$ and $T_{C}$ for $B \parallel a, b$ and $c$ are given by red, blue and green symbols. The blue dashed line represents the function $f(B) = 1 - (B/B_c)^2$, where $B_c$ = 19.6~T.}
\label{figure:phasediagram}
\end{figure}

In Fig.~\ref{figure:phasediagram} we present the superconducting and ferromagnetic phase diagram of UCoGe determined by dilatometry. The $T_{sc}(B)$ and $T_C(B)$ data points are taken from the transverse and longitudinal thermal expansion experiments, respectively. In the case of $B_{c2}$ for $B \parallel b$ we trace the transition points determined by the minimum of $\alpha_b (T)$ (see Fig.~5). For $B \parallel c$ the Curie temperature cannot be identified in field. The magnetic transition smears out rapidly, and the superconducting transition is suppressed near 0.5~T. For $B \parallel a$ the Curie temperature is constant within the error bar at least up to 12~T, while $T_{sc}$ is gradually suppressed, shifting out of the measurement window in fields exceeding 4~T. For $B \parallel b$ $T_C$ shifts towards lower temperatures with increasing field. $B_{c2}$ along the \emph{b}-axis exhibits a remarkable $S$-shape with reinforcement of superconductivity above 6~T. The data in Fig.~\ref{figure:phasediagram} therefore provide bulk evidence for the reinforcement of superconductivity in fields $B \parallel b$. It should be noted that recent thermal conductivity data also show bulk superconductivity up to 15~T for $B \parallel b$~\cite{Wu}.

The field tuning of the Curie point can be fitted using a quadratic function proposed by Mineev~\cite{Mineev2010,Mineev2010PRB}, $T_C (B) = T_C (0)\left[1 - (B/B_c)^2\right]$, assuming a second-order phase transition. Extrapolation of the fit to this expression to $T \rightarrow 0$ indicates the critical field $B_c$ could amount to as much as 19.6~T (see Fig.~\ref{figure:phasediagram}). This value is higher than $ B_c = 16$~T deduced from the transport data~\cite{AokiJPSJ2009}. On the other hand, $T_C$ for $B \parallel b$ is determined from the longitudinal thermal expansion experiment for which precise field-angle tuning was not possible. As it is likely that $T_C (B)$, just like $T_{sc} (B)$, depends strongly on the field-angle, a small misorientation could therefore result in a larger value for $B_c$. We remark that $B_{c2}$ for $B \parallel b$ extracted from the longitudinal thermal expansion (see the inset of Fig.~3) does not show the characteristic $S$-shape seen in Fig.~\ref{figure:phasediagram}, which can be explained by a small misorientation of the $b$-axis with respect to magnetic field in the longitudinal data.

The low-temperature ($T < 0.15$~K) upturn in the $\alpha_b (T)$-data for crystal~2 remains puzzling. A similar upturn is not seen in $\alpha_a (T)$ and $\alpha_c (T)$. Therefore we can safely exclude it is due to an artefact of the experiment, such as an undesired cell-effect (\textit{i.e.} the expansion of the dilatometer itself), or another measurement-related technical problem. Also it was not observed in the $\alpha_b (T)$-data taken on a different crystal~\cite{Gasparini2010}. The possibility that the upturn is a high-temperature tail of a Schottky anomaly due to nuclear magnetic moments can be excluded since the upturn is insensitive to magnetic fields $B \parallel b$ (see Fig.~\ref{figure:lfte2}). Moreover, an analogously anomalous contribution to the specific heat below 0.15~K is not observed (see Fig.~6). We conclude the origin of the upturn in $\alpha_b$ is not understood and further experimental work is needed. The low temperature specific heat data point to a finite $\gamma$-value of 30~mJ/molK$^2$ when $c/T$ as a function of $T$ is linearly extrapolated to $T \rightarrow 0$~\cite{AokiJPSJ2012}. This value is about half of the normal state $\gamma$-value, which could indicate that only one of the spin-split bands takes part in the superconducting condensate~\cite{Mineev2004,Belitz}.

\section{Conclusions}

We have measured the thermal expansion of a single crystal of UCoGe around the ferromagnetic and superconducting transitions in magnetic fields up to 12~T applied along the orthorhombic axes. In order to enable fine-tuning of the magnetic field angle our home-built compact dilatometer was mounted on a piezo-electric rotator. Pronounced steps in the thermal expansion coefficient $\alpha$ were detected at $T_C$ and $T_{sc}$ and their field variation was used to establish the ferromagnetic and superconducting phase diagram. For small fields $B \parallel c$ the ferromagnetic transition becomes a cross-over and superconductivity is rapidly suppressed ($B_{c2} = 0.5$~T when $T \rightarrow 0$). For $B \parallel a,b$ the Curie point and superconductivity persist. With our bulk technique we confirm the $S$-shape of the upper-critical field, $B_{c2}$, for $B \parallel b$ and reinforcement of superconductivity above 6~T. At the same time the Curie point shifts towards lower temperatures. This lends further support to theoretical proposals of spin-fluctuation mediated reinforcement of superconductivity in UCoGe for $B \parallel b$.

\noindent
\textbf{Acknowlegdments}
 This work was funded by the Dutch Foundation for Fundamental Research of Matter (FOM) which is part of the Netherlands Organisation for Scientific Research (NWO). D. Aoki acknowledges support by KAKENHI (15H05884, 15H05882, 15K21732, 16H04006, 15H05745).


\begin{thebibliography}{36}%
\makeatletter
\providecommand \@ifxundefined [1]{%
 \@ifx{#1\undefined}
}%
\providecommand \@ifnum [1]{%
 \ifnum #1\expandafter \@firstoftwo
 \else \expandafter \@secondoftwo
 \fi
}%
\providecommand \@ifx [1]{%
 \ifx #1\expandafter \@firstoftwo
 \else \expandafter \@secondoftwo
 \fi
}%
\providecommand \natexlab [1]{#1}%
\providecommand \enquote  [1]{``#1''}%
\providecommand \bibnamefont  [1]{#1}%
\providecommand \bibfnamefont [1]{#1}%
\providecommand \citenamefont [1]{#1}%
\providecommand \href@noop [0]{\@secondoftwo}%
\providecommand \href [0]{\begingroup \@sanitize@url \@href}%
\providecommand \@href[1]{\@@startlink{#1}\@@href}%
\providecommand \@@href[1]{\endgroup#1\@@endlink}%
\providecommand \@sanitize@url [0]{\catcode `\\12\catcode `\$12\catcode
  `\&12\catcode `\#12\catcode `\^12\catcode `\_12\catcode `\%12\relax}%
\providecommand \@@startlink[1]{}%
\providecommand \@@endlink[0]{}%
\providecommand \url  [0]{\begingroup\@sanitize@url \@url }%
\providecommand \@url [1]{\endgroup\@href {#1}{\urlprefix }}%
\providecommand \urlprefix  [0]{URL }%
\providecommand \Eprint [0]{\href }%
\providecommand \doibase [0]{http://dx.doi.org/}%
\providecommand \selectlanguage [0]{\@gobble}%
\providecommand \bibinfo  [0]{\@secondoftwo}%
\providecommand \bibfield  [0]{\@secondoftwo}%
\providecommand \translation [1]{[#1]}%
\providecommand \BibitemOpen [0]{}%
\providecommand \bibitemStop [0]{}%
\providecommand \bibitemNoStop [0]{.\EOS\space}%
\providecommand \EOS [0]{\spacefactor3000\relax}%
\providecommand \BibitemShut  [1]{\csname bibitem#1\endcsname}%
\let\auto@bib@innerbib\@empty
\bibitem [{\citenamefont {Ginzburg}(1957)}]{Ginzburg1957}%
  \BibitemOpen
  \bibfield  {author} {\bibinfo {author} {\bibfnamefont {V.~L.}\ \bibnamefont
  {Ginzburg}},\ }\href@noop {} {\bibfield  {journal} {\bibinfo  {journal} {Sov.
  Phys. JETP}\ }\textbf {\bibinfo {volume} {4}},\ \bibinfo {pages} {153}
  (\bibinfo {year} {1957})}\BibitemShut {NoStop}%
\bibitem [{\citenamefont {Fay}\ and\ \citenamefont {Appel}(1980)}]{FayAppel}%
  \BibitemOpen
  \bibfield  {author} {\bibinfo {author} {\bibfnamefont {J.}~\bibnamefont
  {Fay}}\ and\ \bibinfo {author} {\bibfnamefont {D.}~\bibnamefont {Appel}},\
  }\href@noop {} {\bibfield  {journal} {\bibinfo  {journal} {Phys. Rev. B}\
  }\textbf {\bibinfo {volume} {22}},\ \bibinfo {pages} {3173} (\bibinfo {year}
  {1980})}\BibitemShut {NoStop}%
\bibitem [{\citenamefont {Saxena}\ \emph {et~al.}(2000)\citenamefont {Saxena},
  \citenamefont {Agarwal}, \citenamefont {Ahilan}, \citenamefont {Grosche},
  \citenamefont {Haselwimmer}, \citenamefont {Steiner}, \citenamefont {Pugh},
  \citenamefont {Walker}, \citenamefont {Julian}, \citenamefont {Monthoux},
  \citenamefont {Lonzarich}, \citenamefont {Huxley}, \citenamefont {Sheikin},
  \citenamefont {Braithwaite},\ and\ \citenamefont {Flouquet}}]{Saxena2000}%
  \BibitemOpen
  \bibfield  {author} {\bibinfo {author} {\bibfnamefont {S.~S.}\ \bibnamefont
  {Saxena}}, \bibinfo {author} {\bibfnamefont {P.}~\bibnamefont {Agarwal}},
  \bibinfo {author} {\bibfnamefont {K.}~\bibnamefont {Ahilan}}, \bibinfo
  {author} {\bibfnamefont {F.~M.}\ \bibnamefont {Grosche}}, \bibinfo {author}
  {\bibfnamefont {R.~K.~W.}\ \bibnamefont {Haselwimmer}}, \bibinfo {author}
  {\bibfnamefont {M.~J.}\ \bibnamefont {Steiner}}, \bibinfo {author}
  {\bibfnamefont {E.}~\bibnamefont {Pugh}}, \bibinfo {author} {\bibfnamefont
  {I.~R.}\ \bibnamefont {Walker}}, \bibinfo {author} {\bibfnamefont {S.~R.}\
  \bibnamefont {Julian}}, \bibinfo {author} {\bibfnamefont {P.}~\bibnamefont
  {Monthoux}}, \bibinfo {author} {\bibfnamefont {G.~G.}\ \bibnamefont
  {Lonzarich}}, \bibinfo {author} {\bibfnamefont {A.}~\bibnamefont {Huxley}},
  \bibinfo {author} {\bibfnamefont {I.}~\bibnamefont {Sheikin}}, \bibinfo
  {author} {\bibfnamefont {D.}~\bibnamefont {Braithwaite}}, \ and\ \bibinfo
  {author} {\bibfnamefont {J.}~\bibnamefont {Flouquet}},\ }\href@noop {}
  {\bibfield  {journal} {\bibinfo  {journal} {Nature (London)}\ }\textbf
  {\bibinfo {volume} {406}},\ \bibinfo {pages} {587} (\bibinfo {year}
  {2000})}\BibitemShut {NoStop}%
\bibitem [{\citenamefont {Aoki}\ \emph {et~al.}(2001)\citenamefont {Aoki},
  \citenamefont {Huxley}, \citenamefont {Ressouche}, \citenamefont
  {Braithwaite}, \citenamefont {Flouquet}, \citenamefont {Brison},
  \citenamefont {Lhotel},\ and\ \citenamefont {Paulsen}}]{Aoki2001}%
  \BibitemOpen
  \bibfield  {author} {\bibinfo {author} {\bibfnamefont {D.}~\bibnamefont
  {Aoki}}, \bibinfo {author} {\bibfnamefont {A.}~\bibnamefont {Huxley}},
  \bibinfo {author} {\bibfnamefont {E.}~\bibnamefont {Ressouche}}, \bibinfo
  {author} {\bibfnamefont {D.}~\bibnamefont {Braithwaite}}, \bibinfo {author}
  {\bibfnamefont {J.}~\bibnamefont {Flouquet}}, \bibinfo {author}
  {\bibfnamefont {J.~P.}\ \bibnamefont {Brison}}, \bibinfo {author}
  {\bibfnamefont {E.}~\bibnamefont {Lhotel}}, \ and\ \bibinfo {author}
  {\bibfnamefont {C.}~\bibnamefont {Paulsen}},\ }\href@noop {} {\bibfield
  {journal} {\bibinfo  {journal} {Nature (London)}\ }\textbf {\bibinfo {volume}
  {413}},\ \bibinfo {pages} {613} (\bibinfo {year} {2001})}\BibitemShut
  {NoStop}%
\bibitem [{\citenamefont {Huy}\ \emph {et~al.}(2007)\citenamefont {Huy},
  \citenamefont {Gasparini}, \citenamefont {de~Nijs}, \citenamefont {Huang},
  \citenamefont {Klaasse}, \citenamefont {Gortenmulder}, \citenamefont
  {de~Visser}, \citenamefont {Hamann}, \citenamefont {G{\"o}rlach},\ and\
  \citenamefont {von L{\"o}hneysen}}]{Huy2007}%
  \BibitemOpen
  \bibfield  {author} {\bibinfo {author} {\bibfnamefont {N.~T.}\ \bibnamefont
  {Huy}}, \bibinfo {author} {\bibfnamefont {A.}~\bibnamefont {Gasparini}},
  \bibinfo {author} {\bibfnamefont {D.~E.}\ \bibnamefont {de~Nijs}}, \bibinfo
  {author} {\bibfnamefont {Y.}~\bibnamefont {Huang}}, \bibinfo {author}
  {\bibfnamefont {J.~C.~P.}\ \bibnamefont {Klaasse}}, \bibinfo {author}
  {\bibfnamefont {T.}~\bibnamefont {Gortenmulder}}, \bibinfo {author}
  {\bibfnamefont {A.}~\bibnamefont {de~Visser}}, \bibinfo {author}
  {\bibfnamefont {A.}~\bibnamefont {Hamann}}, \bibinfo {author} {\bibfnamefont
  {T.}~\bibnamefont {G{\"o}rlach}}, \ and\ \bibinfo {author} {\bibfnamefont
  {H.}~\bibnamefont {von L{\"o}hneysen}},\ }\href@noop {} {\bibfield  {journal}
  {\bibinfo  {journal} {Phys. Rev. Lett.}\ }\textbf {\bibinfo {volume} {99}},\
  \bibinfo {pages} {067006} (\bibinfo {year} {2007})}\BibitemShut {NoStop}%
\bibitem [{\citenamefont {Aoki}\ and\ \citenamefont
  {Flouquet}(2014)}]{AokiJPSJ2014}%
  \BibitemOpen
  \bibfield  {author} {\bibinfo {author} {\bibfnamefont {D.}~\bibnamefont
  {Aoki}}\ and\ \bibinfo {author} {\bibfnamefont {J.}~\bibnamefont
  {Flouquet}},\ }\href {\doibase 10.7566/JPSJ.83.061011} {\bibfield  {journal}
  {\bibinfo  {journal} {J. Phys. Soc. Jpn.}\ }\textbf {\bibinfo {volume}
  {83}},\ \bibinfo {pages} {061011} (\bibinfo {year} {2014})}\BibitemShut
  {NoStop}%
\bibitem [{\citenamefont {Huxley}(2015)}]{Huxley2015}%
  \BibitemOpen
  \bibfield  {author} {\bibinfo {author} {\bibfnamefont {A.~D.}\ \bibnamefont
  {Huxley}},\ }\href@noop {} {\bibfield  {journal} {\bibinfo  {journal}
  {Physica C}\ }\textbf {\bibinfo {volume} {514}},\ \bibinfo {pages} {368 }
  (\bibinfo {year} {2015})}\BibitemShut {NoStop}%
\bibitem [{\citenamefont {Canepa}\ \emph {et~al.}(1996)\citenamefont {Canepa},
  \citenamefont {Manfrinetti}, \citenamefont {Pani},\ and\ \citenamefont
  {Palenzona}}]{Canepa}%
  \BibitemOpen
  \bibfield  {author} {\bibinfo {author} {\bibfnamefont {F.}~\bibnamefont
  {Canepa}}, \bibinfo {author} {\bibfnamefont {P.}~\bibnamefont {Manfrinetti}},
  \bibinfo {author} {\bibfnamefont {M.}~\bibnamefont {Pani}}, \ and\ \bibinfo
  {author} {\bibfnamefont {A.}~\bibnamefont {Palenzona}},\ }\href@noop {}
  {\bibfield  {journal} {\bibinfo  {journal} {J. Alloy. Compd.}\ }\textbf
  {\bibinfo {volume} {234}},\ \bibinfo {pages} {225} (\bibinfo {year}
  {1996})}\BibitemShut {NoStop}%
\bibitem [{\citenamefont {de~Visser}\ \emph {et~al.}(2009)\citenamefont
  {de~Visser}, \citenamefont {Huy}, \citenamefont {Gasparini}, \citenamefont
  {de~Nijs}, \citenamefont {Andreica}, \citenamefont {Baines},\ and\
  \citenamefont {Amato}}]{VissermRS}%
  \BibitemOpen
  \bibfield  {author} {\bibinfo {author} {\bibfnamefont {A.}~\bibnamefont
  {de~Visser}}, \bibinfo {author} {\bibfnamefont {N.~T.}\ \bibnamefont {Huy}},
  \bibinfo {author} {\bibfnamefont {A.}~\bibnamefont {Gasparini}}, \bibinfo
  {author} {\bibfnamefont {D.~E.}\ \bibnamefont {de~Nijs}}, \bibinfo {author}
  {\bibfnamefont {D.}~\bibnamefont {Andreica}}, \bibinfo {author}
  {\bibfnamefont {C.}~\bibnamefont {Baines}}, \ and\ \bibinfo {author}
  {\bibfnamefont {A.}~\bibnamefont {Amato}},\ }\href@noop {} {\bibfield
  {journal} {\bibinfo  {journal} {Phys. Rev. Lett.}\ }\textbf {\bibinfo
  {volume} {102}},\ \bibinfo {pages} {167003} (\bibinfo {year}
  {2009})}\BibitemShut {NoStop}%
\bibitem [{\citenamefont {Ohta}\ \emph {et~al.}(2010)\citenamefont {Ohta},
  \citenamefont {Hattori}, \citenamefont {Ishida}, \citenamefont {Nakai},
  \citenamefont {Osaki}, \citenamefont {Deguchi}, \citenamefont {Sato},\ and\
  \citenamefont {Satoh}}]{Ohta2010}%
  \BibitemOpen
  \bibfield  {author} {\bibinfo {author} {\bibfnamefont {T.}~\bibnamefont
  {Ohta}}, \bibinfo {author} {\bibfnamefont {T.}~\bibnamefont {Hattori}},
  \bibinfo {author} {\bibfnamefont {K.}~\bibnamefont {Ishida}}, \bibinfo
  {author} {\bibfnamefont {Y.}~\bibnamefont {Nakai}}, \bibinfo {author}
  {\bibfnamefont {E.}~\bibnamefont {Osaki}}, \bibinfo {author} {\bibfnamefont
  {K.}~\bibnamefont {Deguchi}}, \bibinfo {author} {\bibfnamefont {N.~K.}\
  \bibnamefont {Sato}}, \ and\ \bibinfo {author} {\bibfnamefont
  {I.}~\bibnamefont {Satoh}},\ }\href@noop {} {\bibfield  {journal} {\bibinfo
  {journal} {J. Phys. Soc. Jpn.}\ }\textbf {\bibinfo {volume} {79}},\ \bibinfo
  {pages} {023707} (\bibinfo {year} {2010})}\BibitemShut {NoStop}%
\bibitem [{\citenamefont {Huy}\ \emph {et~al.}(2008)\citenamefont {Huy},
  \citenamefont {de~Nijs}, \citenamefont {Huang},\ and\ \citenamefont
  {de~Visser}}]{Huy2008}%
  \BibitemOpen
  \bibfield  {author} {\bibinfo {author} {\bibfnamefont {N.~T.}\ \bibnamefont
  {Huy}}, \bibinfo {author} {\bibfnamefont {D.~E.}\ \bibnamefont {de~Nijs}},
  \bibinfo {author} {\bibfnamefont {Y.~K.}\ \bibnamefont {Huang}}, \ and\
  \bibinfo {author} {\bibfnamefont {A.}~\bibnamefont {de~Visser}},\ }\href@noop
  {} {\bibfield  {journal} {\bibinfo  {journal} {Phys. Rev. Lett.}\ }\textbf
  {\bibinfo {volume} {100}},\ \bibinfo {pages} {077002} (\bibinfo {year}
  {2008})}\BibitemShut {NoStop}%
\bibitem [{\citenamefont {Aoki}\ \emph {et~al.}(2009)\citenamefont {Aoki},
  \citenamefont {Matsuda}, \citenamefont {Taufour}, \citenamefont {Hassinger},
  \citenamefont {Knebel},\ and\ \citenamefont {Flouquet}}]{AokiJPSJ2009}%
  \BibitemOpen
  \bibfield  {author} {\bibinfo {author} {\bibfnamefont {D.}~\bibnamefont
  {Aoki}}, \bibinfo {author} {\bibfnamefont {T.~D.}\ \bibnamefont {Matsuda}},
  \bibinfo {author} {\bibfnamefont {V.}~\bibnamefont {Taufour}}, \bibinfo
  {author} {\bibfnamefont {E.}~\bibnamefont {Hassinger}}, \bibinfo {author}
  {\bibfnamefont {G.}~\bibnamefont {Knebel}}, \ and\ \bibinfo {author}
  {\bibfnamefont {J.}~\bibnamefont {Flouquet}},\ }\href@noop {} {\bibfield
  {journal} {\bibinfo  {journal} {J. Phys. Soc. Jpn.}\ }\textbf {\bibinfo
  {volume} {78}},\ \bibinfo {pages} {113709} (\bibinfo {year}
  {2009})}\BibitemShut {NoStop}%
\bibitem [{\citenamefont {Ihara}\ \emph {et~al.}(2010)\citenamefont {Ihara},
  \citenamefont {Hattori}, \citenamefont {Ishida}, \citenamefont {Nakai},
  \citenamefont {Osaki}, \citenamefont {Deguchi}, \citenamefont {Sato},\ and\
  \citenamefont {Satoh}}]{Ihara2010}%
  \BibitemOpen
  \bibfield  {author} {\bibinfo {author} {\bibfnamefont {Y.}~\bibnamefont
  {Ihara}}, \bibinfo {author} {\bibfnamefont {T.}~\bibnamefont {Hattori}},
  \bibinfo {author} {\bibfnamefont {K.}~\bibnamefont {Ishida}}, \bibinfo
  {author} {\bibfnamefont {Y.}~\bibnamefont {Nakai}}, \bibinfo {author}
  {\bibfnamefont {E.}~\bibnamefont {Osaki}}, \bibinfo {author} {\bibfnamefont
  {K.}~\bibnamefont {Deguchi}}, \bibinfo {author} {\bibfnamefont {N.~K.}\
  \bibnamefont {Sato}}, \ and\ \bibinfo {author} {\bibfnamefont
  {I.}~\bibnamefont {Satoh}},\ }\href@noop {} {\bibfield  {journal} {\bibinfo
  {journal} {Phys. Rev. Lett.}\ }\textbf {\bibinfo {volume} {105}},\ \bibinfo
  {pages} {206403} (\bibinfo {year} {2010})}\BibitemShut {NoStop}%
\bibitem [{\citenamefont {Hattori}\ \emph {et~al.}(2012)\citenamefont
  {Hattori}, \citenamefont {Ihara}, \citenamefont {Nakai}, \citenamefont
  {Ishida}, \citenamefont {Tada}, \citenamefont {Fujimoto}, \citenamefont
  {Kawakami}, \citenamefont {Osaki}, \citenamefont {Deguchi}, \citenamefont
  {Sato},\ and\ \citenamefont {Satoh}}]{Hattori2012}%
  \BibitemOpen
  \bibfield  {author} {\bibinfo {author} {\bibfnamefont {T.}~\bibnamefont
  {Hattori}}, \bibinfo {author} {\bibfnamefont {Y.}~\bibnamefont {Ihara}},
  \bibinfo {author} {\bibfnamefont {Y.}~\bibnamefont {Nakai}}, \bibinfo
  {author} {\bibfnamefont {K.}~\bibnamefont {Ishida}}, \bibinfo {author}
  {\bibfnamefont {Y.}~\bibnamefont {Tada}}, \bibinfo {author} {\bibfnamefont
  {S.}~\bibnamefont {Fujimoto}}, \bibinfo {author} {\bibfnamefont
  {N.}~\bibnamefont {Kawakami}}, \bibinfo {author} {\bibfnamefont
  {E.}~\bibnamefont {Osaki}}, \bibinfo {author} {\bibfnamefont
  {K.}~\bibnamefont {Deguchi}}, \bibinfo {author} {\bibfnamefont {N.~K.}\
  \bibnamefont {Sato}}, \ and\ \bibinfo {author} {\bibfnamefont
  {I.}~\bibnamefont {Satoh}},\ }\href@noop {} {\bibfield  {journal} {\bibinfo
  {journal} {Phys. Rev. Lett.}\ }\textbf {\bibinfo {volume} {108}},\ \bibinfo
  {pages} {066403} (\bibinfo {year} {2012})}\BibitemShut {NoStop}%
\bibitem [{\citenamefont {Stock}\ \emph {et~al.}(2011)\citenamefont {Stock},
  \citenamefont {Sokolov}, \citenamefont {Bourges}, \citenamefont {Tobash},
  \citenamefont {Gorfyk}, \citenamefont {Ronning}, \citenamefont {Bauer},
  \citenamefont {Rule},\ and\ \citenamefont {Huxley}}]{Stock2011}%
  \BibitemOpen
  \bibfield  {author} {\bibinfo {author} {\bibfnamefont {C.}~\bibnamefont
  {Stock}}, \bibinfo {author} {\bibfnamefont {D.~A.}\ \bibnamefont {Sokolov}},
  \bibinfo {author} {\bibfnamefont {P.}~\bibnamefont {Bourges}}, \bibinfo
  {author} {\bibfnamefont {P.~H.}\ \bibnamefont {Tobash}}, \bibinfo {author}
  {\bibfnamefont {K.}~\bibnamefont {Gorfyk}}, \bibinfo {author} {\bibfnamefont
  {F.}~\bibnamefont {Ronning}}, \bibinfo {author} {\bibfnamefont {E.~D.}\
  \bibnamefont {Bauer}}, \bibinfo {author} {\bibfnamefont {K.~C.}\ \bibnamefont
  {Rule}}, \ and\ \bibinfo {author} {\bibfnamefont {A.~D.}\ \bibnamefont
  {Huxley}},\ }\href@noop {} {\bibfield  {journal} {\bibinfo  {journal} {Phys.
  Rev. Lett.}\ }\textbf {\bibinfo {volume} {107}},\ \bibinfo {pages} {187202}
  (\bibinfo {year} {2011})}\BibitemShut {NoStop}%
\bibitem [{\citenamefont {Mineev}(2014)}]{Mineev2014}%
  \BibitemOpen
  \bibfield  {author} {\bibinfo {author} {\bibfnamefont {V.~P.}\ \bibnamefont
  {Mineev}},\ }\href@noop {} {\bibfield  {journal} {\bibinfo  {journal} {Phys.
  Rev. B}\ }\textbf {\bibinfo {volume} {90}},\ \bibinfo {pages} {064506}
  (\bibinfo {year} {2014})}\BibitemShut {NoStop}%
\bibitem [{\citenamefont {Tada}\ \emph {et~al.}(2016)\citenamefont {Tada},
  \citenamefont {Takayoshi},\ and\ \citenamefont {Fijimoto}}]{Tada2016}%
  \BibitemOpen
  \bibfield  {author} {\bibinfo {author} {\bibfnamefont {Y.}~\bibnamefont
  {Tada}}, \bibinfo {author} {\bibfnamefont {S.}~\bibnamefont {Takayoshi}}, \
  and\ \bibinfo {author} {\bibfnamefont {S.}~\bibnamefont {Fijimoto}},\
  }\href@noop {} {\bibfield  {journal} {\bibinfo  {journal} {Phys. Rev. B}\
  }\textbf {\bibinfo {volume} {93}},\ \bibinfo {pages} {174512} (\bibinfo
  {year} {2016})}\BibitemShut {NoStop}%
\bibitem [{\citenamefont {Huy}\ \emph {et~al.}(2009)\citenamefont {Huy},
  \citenamefont {Huang},\ and\ \citenamefont {de~Visser}}]{Huy2009JMMM}%
  \BibitemOpen
  \bibfield  {author} {\bibinfo {author} {\bibfnamefont {N.~T.}\ \bibnamefont
  {Huy}}, \bibinfo {author} {\bibfnamefont {Y.~K.}\ \bibnamefont {Huang}}, \
  and\ \bibinfo {author} {\bibfnamefont {A.}~\bibnamefont {de~Visser}},\
  }\href@noop {} {\bibfield  {journal} {\bibinfo  {journal} {J. Magn. Magn.
  Mat.}\ }\textbf {\bibinfo {volume} {321}},\ \bibinfo {pages} {2691} (\bibinfo
  {year} {2009})}\BibitemShut {NoStop}%
\bibitem [{\citenamefont {Aoki}\ and\ \citenamefont
  {Flouquet}(2012)}]{AokiJPSJ2012}%
  \BibitemOpen
  \bibfield  {author} {\bibinfo {author} {\bibfnamefont {D.}~\bibnamefont
  {Aoki}}\ and\ \bibinfo {author} {\bibfnamefont {J.}~\bibnamefont
  {Flouquet}},\ }\href@noop {} {\bibfield  {journal} {\bibinfo  {journal} {J.
  Phys. Soc. Jpn.}\ }\textbf {\bibinfo {volume} {81}},\ \bibinfo {pages}
  {011003} (\bibinfo {year} {2012})}\BibitemShut {NoStop}%
\bibitem [{\citenamefont {Nikitin}(2017)}]{Nikitinthesis}%
  \BibitemOpen
  \bibfield  {author} {\bibinfo {author} {\bibfnamefont {A.~M.}\ \bibnamefont
  {Nikitin}},\ }\href@noop {} {}PhD thesis, University of Amsterdam\ (\bibinfo
  {year} {2017})\BibitemShut {NoStop}%
\bibitem [{\citenamefont {Schmiedeshoff}\ \emph {et~al.}(2006)\citenamefont
  {Schmiedeshoff}, \citenamefont {Lounsbury}, \citenamefont {Luna},
  \citenamefont {Tracy}, \citenamefont {Schramm}, \citenamefont {Tozer},
  \citenamefont {Correa}, \citenamefont {Hannahs}, \citenamefont {Murphy},
  \citenamefont {Palm}, \citenamefont {Lacerda}, \citenamefont {Bud'ko},
  \citenamefont {Canfield}, \citenamefont {Smith}, \citenamefont {Lashley},\
  and\ \citenamefont {Cooley}}]{Schmiedeshoff2006}%
  \BibitemOpen
  \bibfield  {author} {\bibinfo {author} {\bibfnamefont {G.~M.}\ \bibnamefont
  {Schmiedeshoff}}, \bibinfo {author} {\bibfnamefont {A.~W.}\ \bibnamefont
  {Lounsbury}}, \bibinfo {author} {\bibfnamefont {D.~J.}\ \bibnamefont {Luna}},
  \bibinfo {author} {\bibfnamefont {S.~J.}\ \bibnamefont {Tracy}}, \bibinfo
  {author} {\bibfnamefont {A.~J.}\ \bibnamefont {Schramm}}, \bibinfo {author}
  {\bibfnamefont {S.~W.}\ \bibnamefont {Tozer}}, \bibinfo {author}
  {\bibfnamefont {V.~F.}\ \bibnamefont {Correa}}, \bibinfo {author}
  {\bibfnamefont {S.~T.}\ \bibnamefont {Hannahs}}, \bibinfo {author}
  {\bibfnamefont {T.~P.}\ \bibnamefont {Murphy}}, \bibinfo {author}
  {\bibfnamefont {E.~C.}\ \bibnamefont {Palm}}, \bibinfo {author}
  {\bibfnamefont {A.~H.}\ \bibnamefont {Lacerda}}, \bibinfo {author}
  {\bibfnamefont {S.~L.}\ \bibnamefont {Bud'ko}}, \bibinfo {author}
  {\bibfnamefont {P.~C.}\ \bibnamefont {Canfield}}, \bibinfo {author}
  {\bibfnamefont {J.~L.}\ \bibnamefont {Smith}}, \bibinfo {author}
  {\bibfnamefont {J.~C.}\ \bibnamefont {Lashley}}, \ and\ \bibinfo {author}
  {\bibfnamefont {J.~C.}\ \bibnamefont {Cooley}},\ }\href@noop {} {\bibfield
  {journal} {\bibinfo  {journal} {Rev. Sci. Instrum.}\ }\textbf {\bibinfo
  {volume} {77}},\ \bibinfo {pages} {123907} (\bibinfo {year}
  {2006})}\BibitemShut {NoStop}%
\bibitem [{\citenamefont {Gasparini}\ \emph {et~al.}(2010)\citenamefont
  {Gasparini}, \citenamefont {Huang}, \citenamefont {Hartbaum}, \citenamefont
  {v.~L{\"o}hneysen},\ and\ \citenamefont {de~Visser}}]{Gasparini2010}%
  \BibitemOpen
  \bibfield  {author} {\bibinfo {author} {\bibfnamefont {A.}~\bibnamefont
  {Gasparini}}, \bibinfo {author} {\bibfnamefont {Y.~K.}\ \bibnamefont
  {Huang}}, \bibinfo {author} {\bibfnamefont {J.}~\bibnamefont {Hartbaum}},
  \bibinfo {author} {\bibfnamefont {H.}~\bibnamefont {v.~L{\"o}hneysen}}, \
  and\ \bibinfo {author} {\bibfnamefont {A.}~\bibnamefont {de~Visser}},\
  }\href@noop {} {\bibfield  {journal} {\bibinfo  {journal} {Phys. Rev. B}\
  }\textbf {\bibinfo {volume} {82}},\ \bibinfo {pages} {052502} (\bibinfo
  {year} {2010})}\BibitemShut {NoStop}%
\bibitem [{\citenamefont {Slooten}\ \emph {et~al.}(2009)\citenamefont
  {Slooten}, \citenamefont {Naka}, \citenamefont {Gasparini}, \citenamefont
  {Huang},\ and\ \citenamefont {de~Visser}}]{Slooten2009}%
  \BibitemOpen
  \bibfield  {author} {\bibinfo {author} {\bibfnamefont {E.}~\bibnamefont
  {Slooten}}, \bibinfo {author} {\bibfnamefont {T.}~\bibnamefont {Naka}},
  \bibinfo {author} {\bibfnamefont {A.}~\bibnamefont {Gasparini}}, \bibinfo
  {author} {\bibfnamefont {Y.~K.}\ \bibnamefont {Huang}}, \ and\ \bibinfo
  {author} {\bibfnamefont {A.}~\bibnamefont {de~Visser}},\ }\href@noop {}
  {\bibfield  {journal} {\bibinfo  {journal} {Phys. Rev. Lett.}\ }\textbf
  {\bibinfo {volume} {103}},\ \bibinfo {pages} {097003} (\bibinfo {year}
  {2009})}\BibitemShut {NoStop}%
\bibitem [{\citenamefont {Hassinger}\ \emph {et~al.}(2008)\citenamefont
  {Hassinger}, \citenamefont {Aoki}, \citenamefont {Knebel},\ and\
  \citenamefont {Flouquet}}]{Hassinger2008}%
  \BibitemOpen
  \bibfield  {author} {\bibinfo {author} {\bibfnamefont {E.}~\bibnamefont
  {Hassinger}}, \bibinfo {author} {\bibfnamefont {D.}~\bibnamefont {Aoki}},
  \bibinfo {author} {\bibfnamefont {G.}~\bibnamefont {Knebel}}, \ and\ \bibinfo
  {author} {\bibfnamefont {J.}~\bibnamefont {Flouquet}},\ }\href@noop {}
  {\bibfield  {journal} {\bibinfo  {journal} {J. Phys. Soc. Jpn.}\ }\textbf
  {\bibinfo {volume} {77}},\ \bibinfo {pages} {073703} (\bibinfo {year}
  {2008})}\BibitemShut {NoStop}%
\bibitem [{\citenamefont {Bastien}\ \emph {et~al.}(2016)\citenamefont
  {Bastien}, \citenamefont {Braithwaite}, \citenamefont {Aoki}, \citenamefont
  {Knebel},\ and\ \citenamefont {Flouquet}}]{BastienPRB16}%
  \BibitemOpen
  \bibfield  {author} {\bibinfo {author} {\bibfnamefont {G.}~\bibnamefont
  {Bastien}}, \bibinfo {author} {\bibfnamefont {D.}~\bibnamefont
  {Braithwaite}}, \bibinfo {author} {\bibfnamefont {D.}~\bibnamefont {Aoki}},
  \bibinfo {author} {\bibfnamefont {G.}~\bibnamefont {Knebel}}, \ and\ \bibinfo
  {author} {\bibfnamefont {J.}~\bibnamefont {Flouquet}},\ }\href@noop {}
  {\bibfield  {journal} {\bibinfo  {journal} {Phys. Rev. B}\ }\textbf {\bibinfo
  {volume} {94}},\ \bibinfo {pages} {125110} (\bibinfo {year}
  {2016})}\BibitemShut {NoStop}%
\bibitem [{\citenamefont {van Dijk}\ \emph {et~al.}(1995)\citenamefont {van
  Dijk}, \citenamefont {de~Visser}, \citenamefont {Franse},\ and\ \citenamefont
  {Menovsky}}]{vanDijk1995}%
  \BibitemOpen
  \bibfield  {author} {\bibinfo {author} {\bibfnamefont {N.~H.}\ \bibnamefont
  {van Dijk}}, \bibinfo {author} {\bibfnamefont {A.}~\bibnamefont {de~Visser}},
  \bibinfo {author} {\bibfnamefont {J.~J.~M.}\ \bibnamefont {Franse}}, \ and\
  \bibinfo {author} {\bibfnamefont {A.~A.}\ \bibnamefont {Menovsky}},\
  }\href@noop {} {\bibfield  {journal} {\bibinfo  {journal} {Phys. Rev. B}\
  }\textbf {\bibinfo {volume} {51}},\ \bibinfo {pages} {12665} (\bibinfo {year}
  {1995})}\BibitemShut {NoStop}%
\bibitem [{\citenamefont {van Dijk}\ \emph {et~al.}(1993)\citenamefont {van
  Dijk}, \citenamefont {de~Visser}, \citenamefont {Franse},\ and\ \citenamefont
  {Taillefer}}]{vanDijk1993}%
  \BibitemOpen
  \bibfield  {author} {\bibinfo {author} {\bibfnamefont {N.~H.}\ \bibnamefont
  {van Dijk}}, \bibinfo {author} {\bibfnamefont {A.}~\bibnamefont {de~Visser}},
  \bibinfo {author} {\bibfnamefont {J.~J.~M.}\ \bibnamefont {Franse}}, \ and\
  \bibinfo {author} {\bibfnamefont {L.}~\bibnamefont {Taillefer}},\ }\href@noop
  {} {\bibfield  {journal} {\bibinfo  {journal} {J. Low Temp. Phys.}\ }\textbf
  {\bibinfo {volume} {93}},\ \bibinfo {pages} {101} (\bibinfo {year}
  {1993})}\BibitemShut {NoStop}%
\bibitem [{\citenamefont {Adamska}\ \emph {et~al.}(2010)\citenamefont
  {Adamska}, \citenamefont {Havela}, \citenamefont {Surble}, \citenamefont
  {Heathman}, \citenamefont {Posp\'i\v{s}il},\ and\ \citenamefont
  {Dani\v{s}}}]{Adamska2010}%
  \BibitemOpen
  \bibfield  {author} {\bibinfo {author} {\bibfnamefont {A.~M.}\ \bibnamefont
  {Adamska}}, \bibinfo {author} {\bibfnamefont {L.}~\bibnamefont {Havela}},
  \bibinfo {author} {\bibfnamefont {S.}~\bibnamefont {Surble}}, \bibinfo
  {author} {\bibfnamefont {S.}~\bibnamefont {Heathman}}, \bibinfo {author}
  {\bibfnamefont {J.}~\bibnamefont {Posp\'i\v{s}il}}, \ and\ \bibinfo {author}
  {\bibfnamefont {S.}~\bibnamefont {Dani\v{s}}},\ }\href@noop {} {\bibfield
  {journal} {\bibinfo  {journal} {J. Phys.: Condens. Matter}\ }\textbf
  {\bibinfo {volume} {22}},\ \bibinfo {pages} {275603} (\bibinfo {year}
  {2010})}\BibitemShut {NoStop}%
\bibitem [{\citenamefont {Ma\v{s}kov\'a}\ \emph {et~al.}(2012)\citenamefont
  {Ma\v{s}kov\'a}, \citenamefont {Adamska}, \citenamefont {Havela},
  \citenamefont {Kim-Ngan}, \citenamefont {Przewo\'znik}, \citenamefont
  {Dani\v{s}}, \citenamefont {Kothapalli}, \citenamefont {Kolomiets},
  \citenamefont {Heathman}, \citenamefont {Nakotte},\ and\ \citenamefont
  {Bordallo}}]{Maskova2012}%
  \BibitemOpen
  \bibfield  {author} {\bibinfo {author} {\bibfnamefont {S.}~\bibnamefont
  {Ma\v{s}kov\'a}}, \bibinfo {author} {\bibfnamefont {A.~M.}\ \bibnamefont
  {Adamska}}, \bibinfo {author} {\bibfnamefont {L.}~\bibnamefont {Havela}},
  \bibinfo {author} {\bibfnamefont {N.~T.~H.}\ \bibnamefont {Kim-Ngan}},
  \bibinfo {author} {\bibfnamefont {J.}~\bibnamefont {Przewo\'znik}}, \bibinfo
  {author} {\bibfnamefont {S.}~\bibnamefont {Dani\v{s}}}, \bibinfo {author}
  {\bibfnamefont {K.}~\bibnamefont {Kothapalli}}, \bibinfo {author}
  {\bibfnamefont {A.~V.}\ \bibnamefont {Kolomiets}}, \bibinfo {author}
  {\bibfnamefont {S.}~\bibnamefont {Heathman}}, \bibinfo {author}
  {\bibfnamefont {H.}~\bibnamefont {Nakotte}}, \ and\ \bibinfo {author}
  {\bibfnamefont {H.}~\bibnamefont {Bordallo}},\ }\href@noop {} {\bibfield
  {journal} {\bibinfo  {journal} {J. Alloys Compd.}\ }\textbf {\bibinfo
  {volume} {522}},\ \bibinfo {pages} {130} (\bibinfo {year}
  {2012})}\BibitemShut {NoStop}%
\bibitem [{\citenamefont {de~Visser}\ \emph {et~al.}(1990)\citenamefont
  {de~Visser}, \citenamefont {Franse}, \citenamefont {Lacerda}, \citenamefont
  {Haen},\ and\ \citenamefont {Flouquet}}]{Visser1990}%
  \BibitemOpen
  \bibfield  {author} {\bibinfo {author} {\bibfnamefont {A.}~\bibnamefont
  {de~Visser}}, \bibinfo {author} {\bibfnamefont {J.}~\bibnamefont {Franse}},
  \bibinfo {author} {\bibfnamefont {A.}~\bibnamefont {Lacerda}}, \bibinfo
  {author} {\bibfnamefont {P.}~\bibnamefont {Haen}}, \ and\ \bibinfo {author}
  {\bibfnamefont {J.}~\bibnamefont {Flouquet}},\ }\href@noop {} {\bibfield
  {journal} {\bibinfo  {journal} {Physica B}\ }\textbf {\bibinfo {volume}
  {163}},\ \bibinfo {pages} {49 } (\bibinfo {year} {1990})}\BibitemShut
  {NoStop}%
\bibitem [{\citenamefont {Aoki}\ \emph {et~al.}(2011)\citenamefont {Aoki},
  \citenamefont {Hardy}, \citenamefont {Miyake}, \citenamefont {Taufour},
  \citenamefont {Matsuda},\ and\ \citenamefont {Flouquet}}]{Aoki2011573}%
  \BibitemOpen
  \bibfield  {author} {\bibinfo {author} {\bibfnamefont {D.}~\bibnamefont
  {Aoki}}, \bibinfo {author} {\bibfnamefont {F.}~\bibnamefont {Hardy}},
  \bibinfo {author} {\bibfnamefont {A.}~\bibnamefont {Miyake}}, \bibinfo
  {author} {\bibfnamefont {V.}~\bibnamefont {Taufour}}, \bibinfo {author}
  {\bibfnamefont {T.~D.}\ \bibnamefont {Matsuda}}, \ and\ \bibinfo {author}
  {\bibfnamefont {J.}~\bibnamefont {Flouquet}},\ }\href@noop {} {\bibfield
  {journal} {\bibinfo  {journal} {C. R. Physique}\ }\textbf {\bibinfo {volume}
  {12}},\ \bibinfo {pages} {573 } (\bibinfo {year} {2011})}\BibitemShut
  {NoStop}%
\bibitem [{\citenamefont {Wu~$et$ $al.$}()}]{Wu}%
  \BibitemOpen
  \bibfield  {author} {\bibinfo {author} {\bibfnamefont {B.}~\bibnamefont
  {Wu~$et$ $al.$}},\ }\href@noop {} {\bibinfo  {journal} {to be published}\
  }\BibitemShut {NoStop}%
\bibitem [{\citenamefont {Mineev}(2010{\natexlab{a}})}]{Mineev2010}%
  \BibitemOpen
\bibfield  {journal} {  }\bibfield  {author} {\bibinfo {author} {\bibfnamefont
  {V.~P.}\ \bibnamefont {Mineev}},\ }\href@noop {} {\bibfield  {journal}
  {\bibinfo  {journal} {e-print: arXiv:1002.3510v1}\ } (\bibinfo {year}
  {2010}{\natexlab{a}})}\BibitemShut {NoStop}%
\bibitem [{\citenamefont {Mineev}(2010{\natexlab{b}})}]{Mineev2010PRB}%
  \BibitemOpen
  \bibfield  {author} {\bibinfo {author} {\bibfnamefont {V.~P.}\ \bibnamefont
  {Mineev}},\ }\href {\doibase 10.1103/PhysRevB.81.180504} {\bibfield
  {journal} {\bibinfo  {journal} {Phys. Rev. B}\ }\textbf {\bibinfo {volume}
  {81}},\ \bibinfo {pages} {180504} (\bibinfo {year}
  {2010}{\natexlab{b}})}\BibitemShut {NoStop}%
\bibitem [{\citenamefont {Mineev}\ and\ \citenamefont
  {Champel}(2004)}]{Mineev2004}%
  \BibitemOpen
  \bibfield  {author} {\bibinfo {author} {\bibfnamefont {V.~P.}\ \bibnamefont
  {Mineev}}\ and\ \bibinfo {author} {\bibfnamefont {T.}~\bibnamefont
  {Champel}},\ }\href@noop {} {\bibfield  {journal} {\bibinfo  {journal} {Phys.
  Rev. B}\ }\textbf {\bibinfo {volume} {69}},\ \bibinfo {pages} {144521}
  (\bibinfo {year} {2004})}\BibitemShut {NoStop}%
\bibitem [{\citenamefont {Belitz}\ and\ \citenamefont
  {Kirkpatrick}(2004)}]{Belitz}%
  \BibitemOpen
  \bibfield  {author} {\bibinfo {author} {\bibfnamefont {D.}~\bibnamefont
  {Belitz}}\ and\ \bibinfo {author} {\bibfnamefont {T.~R.}\ \bibnamefont
  {Kirkpatrick}},\ }\href@noop {} {\bibfield  {journal} {\bibinfo  {journal}
  {Phys. Rev. B}\ }\textbf {\bibinfo {volume} {69}},\ \bibinfo {pages} {184502}
  (\bibinfo {year} {2004})}\BibitemShut {NoStop}%
\end{thebibliography}


%

\end{document}